\documentclass[prb,aps,twocolumn,floats]{revtex4}
\usepackage{graphicx}

\renewcommand{\vec}[1]{{\bf #1}}
\newcommand{\nin}{\noindent}
\newcommand{\be}{\begin{equation}}
\newcommand{\ee}{\end{equation}}
\newcommand{\bea}{\begin{eqnarray}}
\newcommand{\eea}{\end{eqnarray}}
\newcommand{\lb}{\left[}
\newcommand{\rb}{\right]}
\newcommand{\lp}{\left(}
\newcommand{\rp}{\right)}
\newcommand{\lf}{\left\{}
\newcommand{\rf}{\right\}}

\renewcommand{\r}{{\bf r}}
\renewcommand{\k}{{\bf k}}

\renewcommand{\H}{{\cal H}}

\newcommand{\A}{{\cal A}}

\begin{document}
%\twocolumn[\hsize\textwidth\columnwidth\hsize\csname @twocolumnfalse\endcsname
\title{Supersymmetry in carbon nanotubes in a transverse magnetic field}

\author{H.-W. Lee}
%\affiliation{School of Physics, Korea Institute for Advanced Study,
%207-43 Cheongryangri-dong, Dongaemun-gu, Seoul 130-012, Korea}
%\affiliation{Physics Research Division, Seoul National University,
%Seoul 151-747, Korea} 
\affiliation{Department of Physics, Pohang
University of Science and Technology, Pohang, Kyungbuk 790-784,
Korea}
\author{Dmitry S. Novikov}
\affiliation{Department of Physics,
Center for Materials Sciences \& Engineering,  \\
Massachusetts Institute of Technology, Cambridge, Massachusetts 02139,
USA}
\date{\today}

\begin{abstract}
Electron properties of carbon nanotubes in a transverse
magnetic field are studied using
a model of a massless Dirac particle on a cylinder.
The problem possesses supersymmetry which protects low-energy states and
ensures stability of the metallic behavior in arbitrarily large fields.
In metallic tubes we find suppression of the Fermi velocity 
at half-filling and enhancement of the density of states.
In semiconducting tubes the energy gap is suppressed.
These features qualitatively persist (although to a smaller degree)
in the presence of electron interactions.
The possibilities of experimental observation of these effects are discussed.
%% Low energy effective Hamiltonian for a carbon nanotube (NT)
%% in a transverse magnetic field is supersymmetric.
%% This yields stability of metallic behavior
%% in the arbitrarily large field for a truly metallic (``armchair'') nanotube,
%% and enhances the density of states at half filling.
%% Energy gaps at half filling in semiconducting NTs are suppressed.
%% At a feasible magnetic field such a suppression of a semiconducting NT
%% gap wins over a Zeeman band splitting. We discuss a possibility to observe
%% a semiconducting NT gap suppression in the
%% shift of the van Hove singularities in the density of states, as well
%% as by the activation conductivity measurements.
\end{abstract}

\maketitle
%\begin{multicols}{2}

%%%%%%%%%%%%%%%%%%%%%%%%%%%%%%%%%%%%%%%%%%%%%%%%%%%%%%%%%%%%%%%%
%%%%%%%%%%%%%%%%%%%%%%%%%%%%%%%%%%%%%%%%%%%%%%%%%%%%%%%%%%%%%%%%
\section{Introduction}
\label{sec:intro}

%\nin
The electronic properties of single-walled carbon nanotubes (NT)
vary for tubes with different structure.
Depending on the angle between the tube axis and graphite lattice,
called the NT chiral angle, the tube electron spectrum can be metallic
or semiconducting.~\cite{SaitoDresselhaus}
The semiconducting band gap of a single-walled NT (SWNT) 
is of the order of $1\,{\rm eV}$
and scales inversely with the NT radius $R$.
Also, in many nominally metallic tubes a minigap appears at the
band center due to the intrinsic tube
curvature.~\cite{Hamada92PRL,Kane97PRL,Kleiner01PRB}
These gaps, recently observed experimentally,~\cite{Zhou00PRL,Ouyang01}
have the size of a few tens of millivolts for SWNT's and scale as $1/R^2$
with the NT radius.

Ajiki and Ando have made a remarkable observation \cite{Ajiki93JPSJ}
that in nanotubes
the metallic behavior is fragile: a metallic NT can be easily
turned into a semiconducting one by applying
a relatively weak parallel magnetic field.
Such a field,
by inducing backscattering between right and left electron modes,
opens a minigap at the band center.
This gap, linear in the field, is given by the magnetic flux
scaled by the flux quantum, $\pi R^2 B/\Phi_0$,
times the semiconducting gap size. For $B\simeq 10\,\,{\rm T}$
the gap is of the order of $10\,{\rm meV}$ for typical SWNT radii.
Effects of parallel field on multi-walled NT's have been reported
in Ref.~\cite{exp-B-parallel}.
Electronic properties are also sensitive to mechanical distortion,
such as twisting, bending, or squashing,~\cite{Rochefort,squashing,Figge}
as well as
to external electric fields.~\cite{Zhou01,Novikov02}

Another interesting observation made in Ref.~\cite{Ajiki93JPSJ}
is that a transverse magnetic field affects electron states
in a way completely
opposite to the parallel field effect.
In metallic tubes the
Fermi velocity is suppressed by a transverse field, while the
density of states near the band center is enhanced. At the same time,
in semiconducting tubes the band gap is suppressed.
The goal of the present work is to rationalize these properties
using the concept of
{\it supersymmetry}.~\cite{Cooper95PR} Supersymmetry has
a profound effect on the low-energy properties by protecting the states
at the band center. We derive a supersymmetric Hamiltonian for
and present a simple analytic theory of the above effects.
We find that the metallic behavior is protected by the supersymmetry
for any magnetic field that is applied perpendicularly to the NT and does not
vary along the tube.
For a uniform transverse field, in particular, the
Fermi velocity in metallic tubes is suppressed by the
uniform transverse field as
\be
v(B)=v/I_0(x), \quad x=4\pi R^2B/\Phi_0,
\ee
where $v$ is the Fermi velocity for $B=0$ and
$I_0(x)$ is the modified Bessel function.
The
density of states near the band center is enhanced by the same factor
$I_0(x)$. We also calculate the suppression of the gap
in semiconducting tubes.

The typical field strength required to make these effects
pronounced is quite high. The fields necessary to significantly alter the
electron dispersion can be estimated from
$\pi R^2B \simeq \Phi_0=hc/e$, which gives many tens of tesla
even for the tubes of the largest available radii.
These fields are not hopelessly strong --- they are available,
for example, in pulsed magnetic field sources \cite{Boebinger}
which allow one to reach fields up to $100\,{\rm T}$.~\cite{100tesla}
Because of that and also because
of the novel features arising from
the supersymmetry and Dirac character of low-energy states,
we believe that this problem is sufficiently interesting.

The paper is organized as follows. We first review the basics
of the carbon $\pi$-electron tight-binding band and its relation
with the massless Dirac equation, paying particular attention
to coupling to external fields. Then we present a theory of
a massless Dirac particle on a cylinder in a transverse magnetic field
and calculate the spectrum and density of states. We analyze both
the metallic and semiconducting tubes. Then we briefly discuss
the behavior in extremely high fields, where a connection can be drawn with
the Landau levels and snake states \cite{snake} considered previously
in the context of the quantum Hall effect.

After that we discuss the effects beyond the Dirac
approximation arising from the next order in the gradient expansion of
the tight-binding problem. These effects are small but interesting,
because they violate supersymmetry and lead to minigaps appearing
at the band center. The effects beyond the Dirac model
are controlled by the so-called trigonal warping interaction.
We consider it in the presence of a magnetic field and show that
its effect depends on the chiral angle and, in particular, is absent
for truly metallic armchair nanotubes.
These effects have been discussed in Refs.~\cite{Saito94,Ajiki96JPSJ},
using a combination of numerical
and analytic methods, for zigzag and armchair tubes.
We extend the results of Refs.~\cite{Saito94,Ajiki96JPSJ}
by considering nanotubes with arbitrary chiral angle
and also in the presence of minigaps of other origin.

We also discuss the experimental implications
of this work. The gap suppression in semiconducting NT's
wins over the Zeeman splitting at reasonable fields.
We consider the possibility to observe
the gap suppression in the tunneling density of states
and in the thermally activated transport regime.

Finally, we consider the competition between 
supersymmetry and electron-electron interactions.
We find that although the effect of supersymmetry is reduced by strong 
interactions, the qualitative features of the spectrum 
(suppression of the plasmon velocity and of the semimetallic gap) are
similar to those of the noninteracting case.

%%%%%%%%%%%%%%%%%%%%%%%%%%%%%%%%%%%%%%%%%%%%%%%%%%%%%%%%
\section{Dirac model for the carbon $\pi$ band and
nanotubes in external fields}
\label{sec:carbon-basic}
%%%%%%%%%%%%%%%%%%%%%%%%%%%%%%%%%%%%%%%%%%%%%%%%%%%%%%%%

%\nin
Here we review the basics of the theory of electron states of
the two-dimensional (2D) carbon monolayer, making a connection with the 2D Dirac equation.
This will provide a good starting point for the following discussion
of nanotubes in external fields. We shall start with the tight-binding
description of the carbon $\pi$ band, following the approach
of Ref.~\cite{DiVincenzo84PRB}, recall how the Dirac equation
arises in this system, and then consider electron
coupling to external electromagnetic fields.

The tight-binding
Hamiltonian on a honeycomb lattice of carbon atoms with hopping
amplitude $t$ between
adjacent sites has the form
\begin{eqnarray}
\epsilon \psi(r)=-t\sum\limits_{|r'-r|=a_{\rm cc}}\psi(r')
\ ,
\label{eq:2Dgraphite}
\end{eqnarray}
where $r'$ are the nearest neighbors of the site $r$,
and $a_{\rm cc}$ is interatomic spacing. 
In carbon, $t\approx 3\,{\rm eV}$ and
$a_{\rm cc}=0.1437\,{\rm nm}$.
For simplicity and because the electron spectrum is
$\epsilon\to-\epsilon$ symmetric, from now on we shall ignore 
the minus sign in Eq.~(\ref{eq:2Dgraphite}).

The zero chemical potential in Eq.~(\ref{eq:2Dgraphite}) describes the
half-filled $\pi$ band, i.e., the density of one electron per site.
For an infinite system, the states of the problem (\ref{eq:2Dgraphite})
are plane waves and
the spectrum is given by $\epsilon(k)=\pm t|\sum_i e^{i\vec k\cdot \vec r_i}|$,
where $\vec r_i$ are the nearest-neighbor bond vectors.
This is a spectrum of a semimetal with
the conduction [$\epsilon(k)>0$] and valence [$\epsilon(k)<0$] subbands
touching each other at two points $K$ and $K'$ in the Brillouin zone.

\begin{figure}
\centering
\includegraphics[width=1.6in]{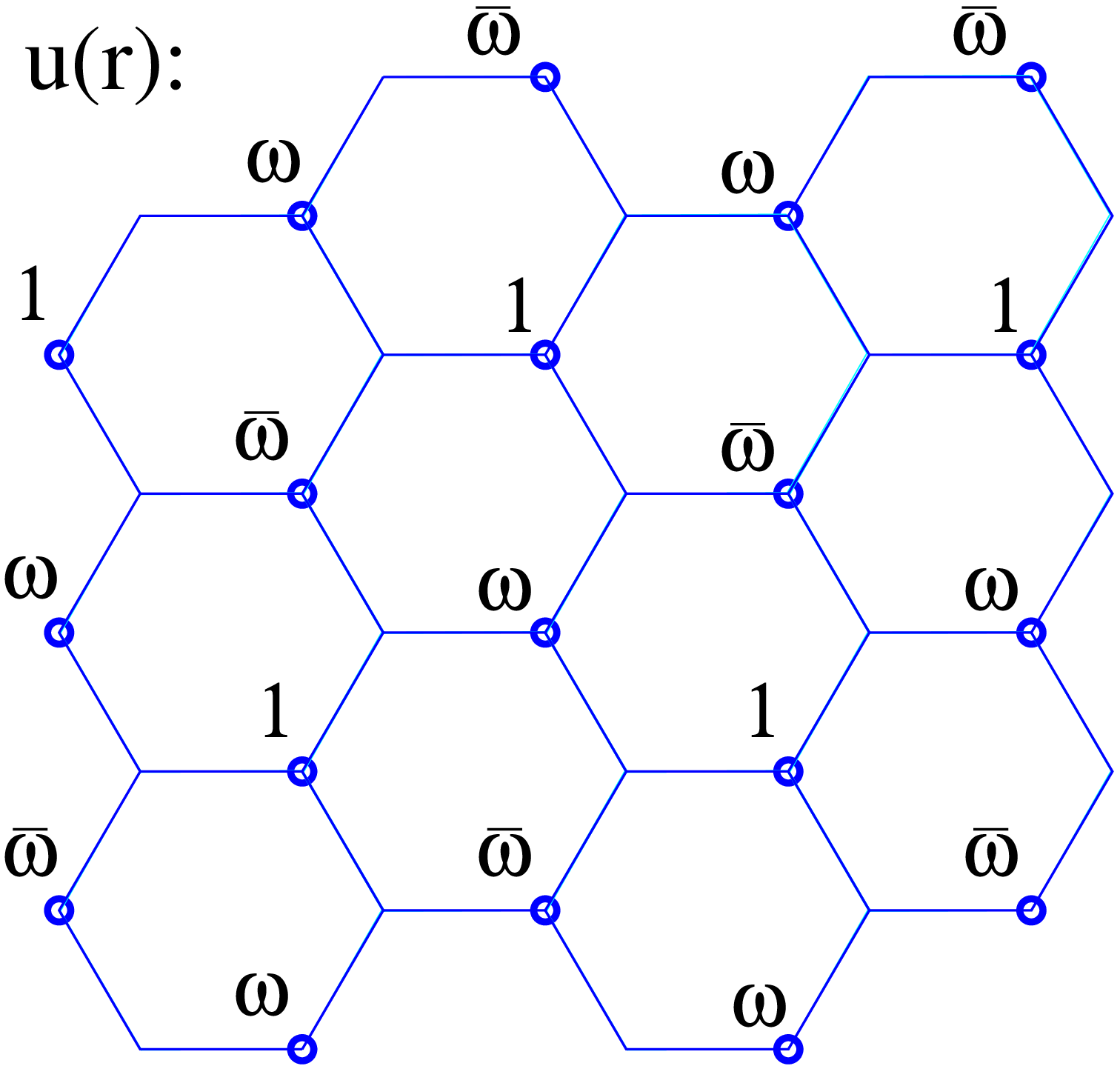}\quad
\includegraphics[width=1.6in]{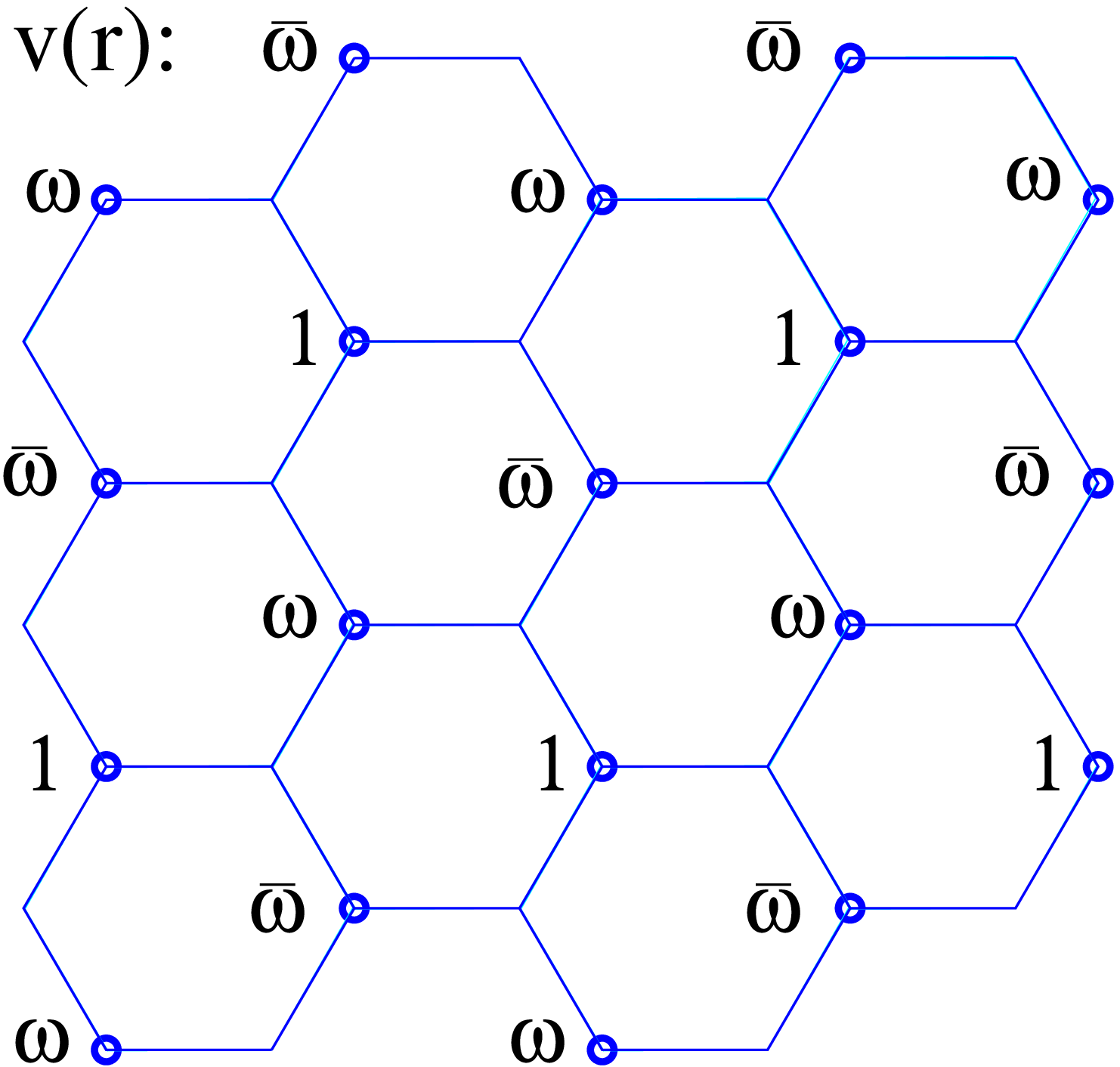}
\vspace{0.5cm}
 \caption[]{Shown are two plane-wave basis states $u(r)$ and $v(r)$
of the problem (\ref{eq:2Dgraphite}) with $\epsilon=0$. Both $u(r)$ and $v(r)$
take the values $1$, $\omega=e^{(2\pi/3)i}$
and $\bar\omega=e^{-(2\pi/3)i}$ on one sublattice and vanish on the
other sublattice of the honeycomb lattice.
The states
$u(r)$ and $v(r)$ have the same quasimomentum and
form a basis
of the Dirac problem (\ref{eq:Dirac-alpha}) at the point $K$.
The independent basis states at the point $K'$
are $\bar u(r)$ and $\bar v(r)$.
    }
 \label{fig:uv-states}
\end{figure}

The tight-binding bandwidth $6t\simeq 18\,{\rm eV}$ is much larger than the
energies of the states close to the band center considered below.
Because of that, it is useful to
project the problem (\ref{eq:2Dgraphite}) onto the subspace of states with
$|\epsilon|\ll t$
and derive an effective low energy Hamiltonian for
such states.
To carry out the projection, we note that there are only four
independent states with $\epsilon=0$. These states
form two complex valued
conjugate pairs which we denote as $u(r)$, $v(r)$ and $\bar u(r)$, $\bar v(r)$.
It is convenient to choose the states $u$ and $v$ to be
zero on one of the two sublattices of the honeycomb lattice.
On the other sublattice each state takes
the values $1$, $\omega=e^{(2\pi/3)i}$ and
$\bar\omega=e^{-(2\pi/3)i}$ (see Fig.~\ref{fig:uv-states}).
The states
$u(r)$ and $v(r)$ have the same quasimomentum
of a value 
\be \label{K0}
K_0={4\pi \over 3\sqrt{3}a_{\rm cc}}  \ ,
\ee
opposite to that of the states $\bar u(r)$ and $\bar v(r)$.
Each pair of states $u(r)$, $v(r)$
and $\bar u(r)$, $\bar v(r)$
forms a basis at the points $K$ and $K'$, respectively.

Projecting the wave function $\psi(r)$ on $u(r)$ and $v(r)$ and, respectively,
on $\bar u(r)$ and $\bar v(r)$ defines Dirac spinor components for
each of the two points $K$ and $K'$. We
focus on the $u$, $v$ pair and write
the states near the point $K$ with small energies
$|\epsilon|\ll t$ as linear combinations
\begin{eqnarray}
\psi(r)=\psi_1(r)u(r)+\psi_2(r)v(r),
\label{eq:psi12}
\end{eqnarray}
with the envelope functions $\psi_{1,2}(r)$ varying on the scale much larger
than the interatomic spacing $a_{\rm cc}$.
By substituting the wave function (\ref{eq:psi12}) in the tight-binding
Hamiltonian (\ref{eq:2Dgraphite}) we have
\begin{eqnarray}
%\label{eq:psi12coupled}
\epsilon\,\psi_1(r)=
t\lf\psi_2(r \!-\! a)+\bar\omega\psi_2(r \!-\! \omega a)+\omega\psi_2(r \!-\! \bar\omega a)\rf,
\label{eq:p1p2}
\\
\epsilon\,\psi_2(r)=
t\lf\psi_1(r \!+\! a)+\omega\psi_1(r \!+\! \omega a)+\bar\omega\psi_1(r \!+\! \bar\omega a)\rf,
\label{eq:p2p1}
\end{eqnarray}
where $a$ is a shorthand notation for $a_{\rm cc}$.
Here the products $za$ with unimodular complex numbers
$z=1,\,\omega,\,\bar\omega$ in the arguments of $\psi_{1,2}$
are understood in terms of 2D rotations of the vector $a{\bf \hat x}$
by ${\rm arg\,}z$.

Expanding slowly varying $\psi_{1,2}(r)$, we obtain
\begin{eqnarray}
\begin{array}{l}
\epsilon\psi_1(r)=
-\hbar v (\partial_x-i\partial_y)\psi_2(r),
%\nonumber
 \\
\epsilon\psi_2(r)=
\hbar v (\partial_x+i\partial_y)\psi_1(r),
\end{array}
\ ,
\label{eq:psi12linearized}
\end{eqnarray}
where $v=\frac32 t\,a_{\rm cc}/\hbar$.
The Hamiltonian (\ref{eq:psi12linearized}) defines massless Dirac fermions
with the linear spectrum $\epsilon(k)=\pm \hbar v|\vec k|$.
In carbon, the velocity $v=8\times 10^7{\rm cm/s}$.
Similar relations hold for the point $K'$.

%%%%%%%%%%%%%%%%%%%%%%%%%%%%%%%%%%%%%%%%%%%%%%%%%%%%%%%%%%%%%%%%%%%%
%%%%%%%%%%%%%%%%%%%%%%%%%%%%%%%%%%%%%%%%%%%%%%%%%%%%%%%%%%%%%%%%%%%%

Equations~(\ref{eq:psi12linearized}) can be cast in the
conventional Dirac form $\epsilon\psi={\cal H}\psi$
with
\be\label{eq:Halpha}
{\cal H}=v\,\vec\alpha\cdot\vec p
=  v \lp \alpha_1p_1+\alpha_2p_2\rp
\ee
for the two-component
wave function $\psi=(\psi_1,\psi_2)^{\rm T}$, with $\alpha_{1,2}$
given by the Pauli matrices:
\begin{eqnarray}
\alpha_1=\sigma_2
\ ,\quad
\alpha_2=-\sigma_1
\ .
\label{eq:Dirac-alpha}
\end{eqnarray}
The Hamiltonian near the $K'$ point can be derived in a similar way.
The result has the form (\ref{eq:Halpha}) with a sign change
in the second term: $\alpha_1=\sigma_2$, $\alpha_2=\sigma_1$.

Below we shall consider
electrons in the presence of external electromagnetic fields.
The minimal form of the coupling to external fields follows from
the gauge invariance:
\begin{eqnarray}
{\cal H}=v\,\vec\alpha\cdot\left(\vec p-{e\over c}\vec A\right)
+e\varphi
\ ,
\label{eq:Dirac-Aphi}
\end{eqnarray}
where $\varphi$ and $\vec A$ are the scalar and vector electromagnetic
potentials. The effect of electron spin, ignored here for simplicity,
can be included in Eq.~(\ref{eq:Dirac-Aphi}) via a Zeeman term.

Equation (\ref{eq:Dirac-Aphi}) describes the lowest-order 
approximation in the gradients of $\psi_{1,2}$ and the potentials
$\varphi$ and $\vec A$. Here we consider the exact tight-binding
equations in the presence of external fields:
\begin{eqnarray}\label{eq:p1p2z}
\epsilon\,\psi_1(r) = t \lp \sum\limits_{z=1,\omega,\bar\omega}
\bar z\,e^{i\gamma_{r,\,r-za}}
\psi_2(r-za)\rp,
\\ \label{eq:p2p1z}
\epsilon\,\psi_2(r) = t \lp \sum\limits_{z=1,\omega,\bar\omega}
z\,e^{i\gamma_{r,\,r+za}}
\psi_1(r+za)\rp,
\end{eqnarray}
where the phases $\gamma_{r,r'}$ are the integrals of the vector
potential along the nearest-neighbor bonds,
\be\label{eq:gamma-rr'}
\gamma_{r,r'}=\frac{2\pi}{\Phi_0}\int\limits_{r'}^r
\vec A(\vec x)\cdot d\vec l \ .
\ee
Equations~(\ref{eq:p1p2z}) and (\ref{eq:p2p1z}) can be used to obtain
the gradient terms of higher order along with the coupling to
external fields.
One can check that expanding the exponents
in Eqs.~(\ref{eq:p1p2z}) and (\ref{eq:p2p1z}) and keeping the
lowest nonvanishing terms gives the Dirac Hamiltonian (\ref{eq:Dirac-Aphi}).
In Sec.~\ref{sec:beyond-dirac} we shall use
Eqs.~(\ref{eq:p1p2z}) and (\ref{eq:p2p1z}) to obtain higher-order corrections
to Eq.~(\ref{eq:Dirac-Aphi}).

To apply the above results to nanotubes, we consider electrons
on a carbon sheet rolled into a cylinder. The transformation of
the tight-binding problem (\ref{eq:2Dgraphite}) to the Dirac problem
(\ref{eq:psi12linearized}) based on the representation (\ref{eq:psi12})
is valid
provided that the cylinder circumference $L=2\pi R$
is much larger than the interatomic spacing $a_{\rm cc}$.
Since for typical NT radii
the ratio $L/a_{\rm cc}$ can be between $10$ and $20$, the
approximation (\ref{eq:psi12}) is entirely adequate.

The NT electron properties, depending on the nanotube
structure, can be either metal like or dielectric like.
Which of these situations takes place depends on the manner the cylinder is
obtained from the carbon monolayer. In the Dirac approach, the condition for
metallic behavior can be formulated directly in terms of
the functions $u(r)$ and $v(r)$:
The nanotube is metallic
{\it if and only if} one can define
on the NT cylinder the two functions $u(r)$ and $v(r)$ according to
Fig.~\ref{fig:uv-states} without running into a mismatch
of the function values upon the cylinder closure.

To demonstrate this, let us suppose that the functions
$u(r)$ and $v(r)$ on the cylinder exist.
Without loss of generality we choose the $x$ axis along the cylinder
and the $y$ axis along the circumference.
The problem (\ref{eq:psi12linearized}) has periodic boundary conditions
in the $y$ direction, and thus the wave functions can be factorized as
$\psi_{1,2}(r)=\psi_{1,2}(x)e^{ik_n y}$, where $k_n=2\pi n/L = n/R$. Then
the dispersion relation for the 1D problems describing
motion along the $x$ axis with fixed $k_n$ is
\begin{eqnarray}
\epsilon_n(k_x)=\pm \hbar v(k_x^2+k_n^2)^{1/2}
\ .
\label{eq:spec-m}
\end{eqnarray}
In this case the subband with $n=0$ has metallic
properties and the subbands with $n\neq0$ are dielectric.

Now let us consider the other possibility when the cylinder
is constructed in such
a way that the functions $u(r)$ and $v(r)$ cannot be defined without
a value mismatch. In this case, upon rolling the carbon sheet
into a cylinder, the sites with different function values
shown in Fig.~\ref{fig:uv-states} are glued together.
However, since all values of the functions $u(r)$ and $v(r)$ are powers
of $\omega=e^{(2\pi/3)i}$, one notes that Eqs.~(\ref{eq:psi12linearized})
can still be used here if they are augmented
with {\it quasiperiodic} boundary conditions,
$\psi_{1,2}(x,y+L)=\omega\psi_{1,2}(x,y)$ or
$\psi_{1,2}(x,y+L)=\bar\omega\psi_{1,2}(x,y)$,
which, combined with the value mismatch of $u(r)$ and $v(r)$, make
$\psi(r)$ single valued.
Factoring the wave function as above, one obtains
1D subbands with the dispersion of the form (\ref{eq:spec-m}),
in this case with $k_n=(n\pm \frac13)/R$. Note that in this case
all spectral branches have dielectric character.

Now we consider a nanotube in the presence of a
{\it parallel} external magnetic field.
In this case, electron properties are described
by the Dirac equation (\ref{eq:Dirac-Aphi}) with $\varphi=0$ and
the vector potential $\vec A$ with just the $y$ component,
$A_y=\Phi/L$, where $\Phi=\pi R^2 B$ is the magnetic flux.
The boundary conditions in the $y$ direction are periodic for
the metallic case and quasiperiodic for the dielectric case.
In the presence of a parallel magnetic field
the problem
remains separable and thus the wave function
can be factorized in just the same way as above. One again finds
1D subbands with the spectrum (\ref{eq:spec-m}), where
\be
%% k_n=\frac{2\pi}{L}(n+\phi)
%% \quad {\rm or}\quad
k_nR=\cases{n+\phi_{\parallel} \ ,\quad & metallic, \cr
n\pm{\textstyle \frac13}+\phi_{\parallel} \  & semiconducting,}
%% \frac{2\pi}{L}(n\pm{\textstyle \frac13}+\phi)
\label{eq:MD-Phi}
\ee
for the metallic and semiconducting NT's, respectively,
with $\phi_{\parallel}=\Phi/\Phi_0$ and $\Phi_0=hc/e$.
Thus in the presence
of a parallel field the gapless $n=0$ branch of the metallic nanotube
spectrum (\ref{eq:spec-m}) acquires a gap.~\cite{Ajiki93JPSJ,Ajiki96JPSJ}
Interestingly, there is no
threshold for this effect, since the gap
forms at arbitrarily weak field. The gap size is
$2\Delta = 2|\phi_{\parallel}| \hbar v/R$.
One notes that the field-induced gap
appears not at the Fermi level but at the center of the electron band.
Thus it affects
the metallic NT properties only for electron
density sufficiently close to half-filling.

%%%%%%%%%%%%%%%%%%%%%%%%%%%%%%%%%%%%%%%%%%%%%%%%%%%%%%%%%%%%%%%%
%%%%%%%%%%%%%%%%%%%%%%%%%%%%%%%%%%%%%%%%%%%%%%%%%%%%%%%%%%%%%%%%
\section{Dirac Equation and Supersymmetry}
\label{sec:dirac}

%\nin
At the energies smaller than the total bandwidth $6t$
(ca. $18\,{\rm eV}$ in carbon) electron states are described
(separately near each of the $K$ and $K'$ points)
by the massless Dirac equation (\ref{eq:Dirac-Aphi}).
For a uniform transverse magnetic field, the
field component normal to the NT surface is
$B_{\perp}(\theta) = B \sin \theta$, where $\theta=y/R$ is the azimuthal angle.
The corresponding vector potential can be chosen along the tube axis $x$,
$\vec A(\vec r)={\bf \hat x} A(\theta)$, where
\be \label{def-A}
{eR\over \hbar c}\, A(\theta) = 2\phi \, \cos \theta \ , \quad
\phi \equiv {\pi R^2B \over \Phi_0} \ .
\ee
In this case the longitudinal momentum $\hbar k$ is conserved
and the states on the NT cylinder have a plane-wave form
$\psi(r)=\psi(x,\theta)=\psi(\theta)e^{ikx}$.
The Dirac Hamiltonian for $\psi(\theta)$ is
\be\label{H-dirac}
\H_D=\Delta_0 \lf i \sigma_1 \partial_\theta +
\lp 
%kR - {eR\over \hbar c}A(\theta)
kR - 2\phi \, \cos \theta
\rp \sigma_2 \rf,
\ee
with $\sigma_{1,2}$ the Pauli matrices and
\be \label{def-delta}
\Delta_0={\hbar v \over R} \ .
\ee
The equations near the $K'$ point have the form (\ref{H-dirac})
with a sign change in the first term, $\sigma_1\to -\sigma_1$.

The eigenvalues of the
operator (\ref{H-dirac}) give the electron dispersion relation $\epsilon(k)$.
We have chosen the dimensionless transverse field parameter $\phi$
in the form (\ref{def-A}), which makes contact
with the parallel field problem,~\cite{Ajiki93JPSJ,Ajiki96JPSJ}
Eq.~(\ref{eq:MD-Phi}).

The NT states are described by quasiperiodic wave functions on the cylinder
($y=R\theta$),
\be\label{def-BC}
\psi(y+L)=e^{2\pi i\delta}\psi(y) \ , \quad L=2\pi R \ ,
\ee
with
\be
\delta =\cases{0 \, ,\quad & metallic, \cr
\pm{\textstyle \frac13} & semiconducting,}
\label{eq:delta}
\ee
We consider the problem (\ref{H-dirac}) with an arbitrary phase $\delta$
in the boundary conditions (\ref{def-BC}). This will permit us to generalize
the results to the cases of metallic NT's with a minigap induced by
curvature~\cite{k-shift} or in the presence of a parallel magnetic field.
These problems can be described using the boundary conditions (\ref{def-BC}) with
$\delta$ slightly shifted away from the ideal values (\ref{eq:delta}).

\begin{figure}[t]
\includegraphics[width=3.5in,height=2.5in]{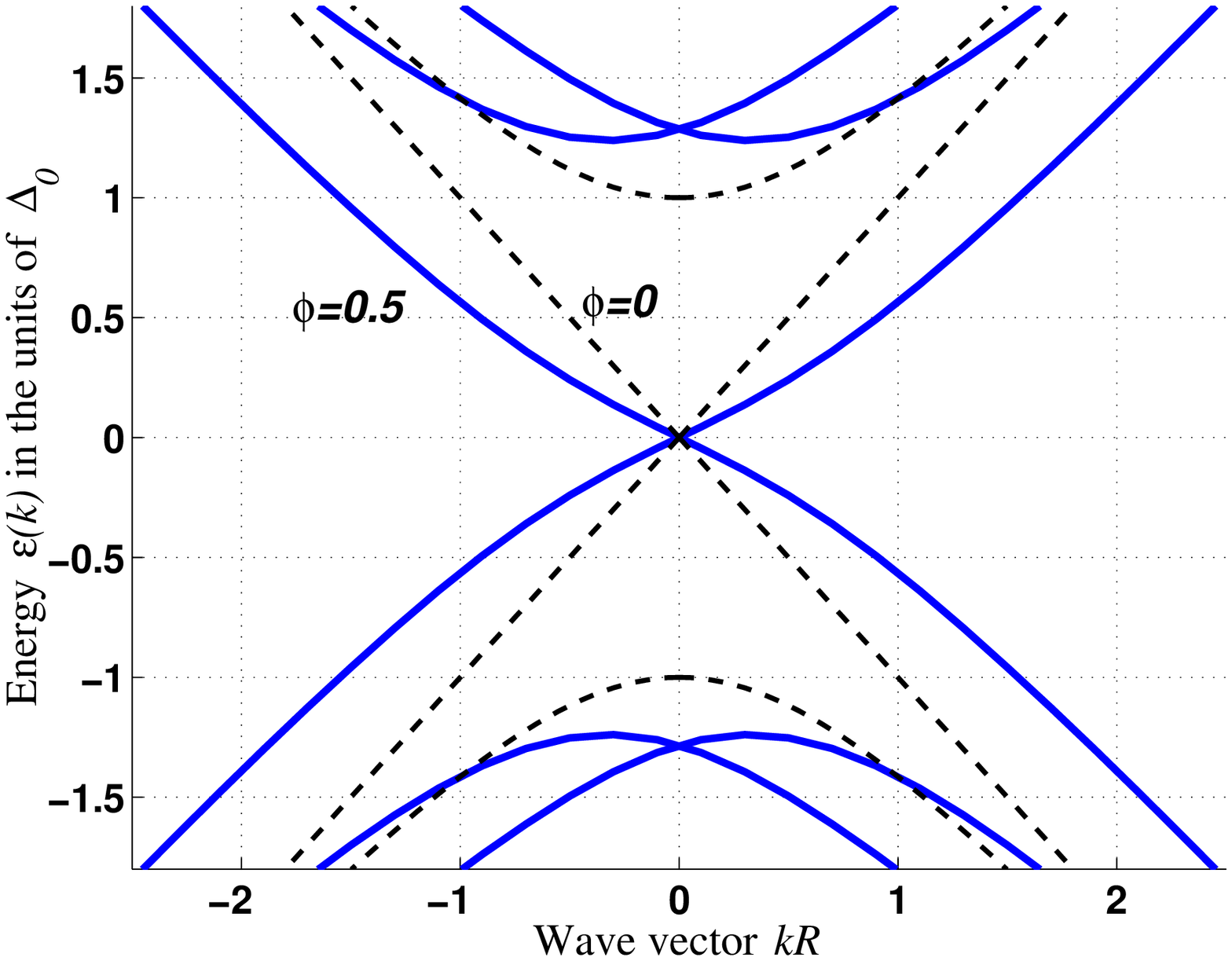}
\vspace{0.1cm}
\includegraphics[width=3.5in,height=2.5in]{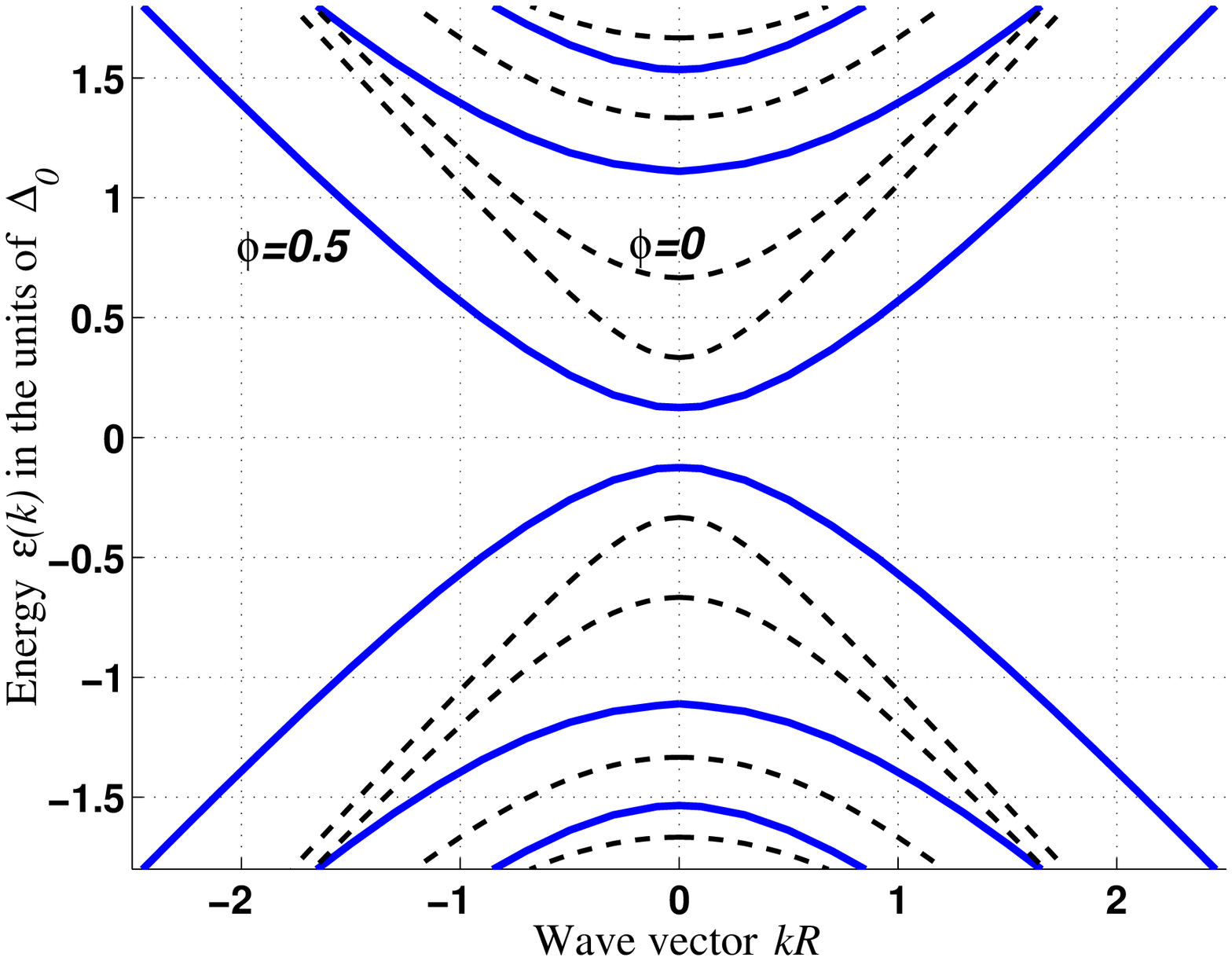}
\vspace{0.1cm} \caption[]{\label{fig:dispersion} Electron
dispersion $\epsilon(k)$ in the presence of a large uniform
transverse magnetic field $\phi=0.5$ (bold lines) and in the absence of
the field (dashed lines). {\it Top:} metallic NT ($\delta=0$). No gap
opens due to supersymmetry, but the velocity is suppressed
according to Eq.~(\ref{g-metallic}). {\it Bottom:} semiconducting
case ($\delta=1/3$). The energy gap is suppressed by the factor
$g_{1/3}(\phi)$; see Eq.~(\ref{g-sc}) and
Fig.~\ref{fig:suppression-factor}. }
\end{figure}

The electron bands $\epsilon(k)$
%% in the entire range of $k$
can be
studied using the transfer matrix.
We integrate the Dirac equation in the interval $0<\theta<2\pi$ and
write
a formal solution $\psi(\theta)$
to the problem $\H_D\psi = \epsilon\psi$
as $\psi(\theta)=S(\theta)\psi(0)$ with
the $2\times2$ matrix
\be \label{def-S} S(\theta)={\cal T}\, {\rm exp} \int_0^\theta
\!\!\! \lf -i\tilde\epsilon\,\sigma_1 + \lp 2\phi\cos\theta' - kR
\rp \sigma_3 \rf d\theta' \ , \ee
where $\tilde\epsilon=\epsilon/\Delta_0$. Here ${\cal T}$ 
stands for operator ordering with respect to
$\theta$. The quasiperiodic boundary condition (\ref{def-BC})
requires
\be
\label{eq:trace-criterion}
{\rm tr\,}\, S_{\theta=2\pi} = 2\cos(2\pi\delta) \ .
\ee
Different energy bands $\epsilon=\epsilon_n(k)$ can be found
numerically as solutions of Eq.~(\ref{eq:trace-criterion}).
The bands obtained in this way
are displayed in Fig.~\ref{fig:dispersion}.

%%%%%%%%%%%%%%%%%%%%%%%%%%%%%%%%%%%%%%%%%%%%%%%%%%%%%%%%%%%%%%%%
%%%%%%%%%%%%%%%%%%%%%%%%%%%%%%%%%%%%%%%%%%%%%%%%%%%%%%%%%%%%%%%%

In the present section we show that supersymmetry allows one to make rather general
statements about the low-energy NT spectrum. 
Originally, supersymmetry was suggested\cite{Golfand}
as a special symmetry between the bosonic and fermionic sectors of relativistic
field theories that protects the zero-energy eigenstate.
Later, the concept of supersymmetry was brought to single-particle quantum 
mechanics.~\cite{Witten} This has yielded the classification of exactly solvable 
potentials using factorization of the Schr\"odinger equation
(see Ref.~\cite{Cooper95PR} for a review).

Below we apply the arguments of supersymmetry to 
a problem of NT electrons in a generic magnetic
field ${\bf B}(\r)$ that (i) is perpendicular to the NT axis and (ii)
does not vary along the NT. In this case the field component
normal to the NT surface is a function only of $y=R\theta$,
$B_{\perp} \equiv B(y)$. Similarly to Eq.~(\ref{def-A}), we choose
the following gauge:
\be
\label{def-A-general}
A_x \equiv A(y) = {\Phi_0\over2\pi}\,
{d\varphi\over dy} \ , \quad A_y=0 \ ,
\ee
where
\be
\label{def-varphi-general} {d^2\varphi\over dy^2} = - {2\pi\over
\Phi_0}\, B(y) \ .
\ee
The function $\varphi(y)$ is uniquely
defined by demanding periodicity 
\be
\varphi(y+L) = \varphi(y) 
\ee
and zero average 
\be
\int_0^L\! dy\, \varphi = 0 \ . 
\ee
The corresponding
Dirac Hamiltonian reads
\be
\label{H-dirac-general}
\tilde\H_D =
\hbar v\lf i \sigma_1\partial_y  + 
\lp k - {d\varphi\over dy} \rp \sigma_2 \rf \ ,
\ee
which reduces to  
Eq.~(\ref{H-dirac}) when $B(y)=B\sin (y/R)$. It is useful to
decompose $\tilde\H_D$ into two pieces, 
\begin{equation}
\tilde\H_D=Q+Q^\dagger,
\end{equation}
where
\begin{equation}
Q\equiv \left( \begin{array}{cc} 0 & 0 \\ \A & 0 \end{array}
\right), \ \ \
Q^\dagger \equiv \left( \begin{array}{cc} 0 & \A^\dagger \\
0 & 0
\end{array} \right)
\label{def-superoperators}
\end{equation}
and 
\begin{eqnarray}
\A & \equiv & \hbar v \left\{ i \partial_y+i\left(k-{d \varphi \over
dy} \right) \right\}, \nonumber \\
\A^\dagger & \equiv & \hbar v \left\{ i \partial_y-i\left(k-{d \varphi
\over dy} \right) \right\}.
\end{eqnarray}
The connection with the supersymmetric quantum mechanics (see 
Chap.~2 in Ref.~\cite{Cooper95PR}) is established by
constructing a supersymmetric Hamiltonian
\begin{equation}
\H_{\rm SUSY}\equiv \left( \begin{array}{cc} \A^\dagger \A & 0 \\ 0 &
\A \A^\dagger \end{array} \right),
\end{equation}
which, together with $Q$ and $Q^\dagger$, satisfies the superalgebra
{\it sl}(1/1),
\begin{eqnarray}
\matrix{
[\H_{\rm SUSY},Q]=[\H_{\rm SUSY},Q^\dagger]=0, 
\cr 
\{Q,Q^\dagger\}=\H_{\rm SUSY}, \quad \{Q,Q\}=\{Q^\dagger,Q^\dagger\}=0. 
}
\label{susyalgebra}
\end{eqnarray}
Here $\{A,B\}=AB+BA$ stands for anticommutator of operators.
In relativistic field theories, the {\sl supercharges} $Q$ and $Q^\dagger$
transform fermionic and bosonic degrees of freedom into each other.
Although we deal with a single electron NT spectrum, 
one could {\it formally} interpret the upper and lower
components of the wave function $\psi$ as fermionic and bosonic sectors of the
supersymmetric Hamiltonian $\H_{\rm SUSY}$.

One interesting implication of the algebra (\ref{susyalgebra}) 
is that  $\H_{\rm SUSY}$ can be expressed as sums of the square of
Hermitian supercharges, $Q_1$ and $Q_2$,
\begin{equation}
\H_{\rm SUSY}=Q_1^2+Q_2^2, \label{squares}
\end{equation}
where
\begin{equation}
Q_1\equiv {1 \over \sqrt{2}}\left( Q+Q^\dagger \right) \ , 
%={\tilde\H_D \over  \sqrt{2}}, 
\quad Q_2\equiv {i \over \sqrt{2}}\left(
Q-Q^\dagger \right).
\end{equation} From Eq.~(\ref{def-superoperators}), one verifies
$Q_1^2=Q_2^2$ and
\begin{equation}
\H_{\rm SUSY}=2Q_1^2=\tilde\H_D^2. \label{squares2}
\end{equation}
Thus the energy spectra of $\H_{\rm SUSY}$ and $\tilde\H_D$ are
closely related.

Let us now show how supersymmetry protects zero-energy states of 
$\H_{\rm SUSY}$ and $\tilde\H_D$.
For that, we construct zero energy states of $\H_{\rm SUSY}$
that are compatible with the boundary condition~(\ref{def-BC}).
Due to Eq.~(\ref{squares}), any such state $\psi$ 
%of $\H_{\rm SUSY}$ 
satisfies 
\be
Q_1\psi=Q_2\psi=0
\ee
or, equivalently,
\be
Q\psi=Q^\dagger\psi=0 \ . 
\ee
The latter equation has two independent solutions
\begin{equation}
\label{indep-solns}
\psi_1 =  e^{-ky+\varphi(y)} \lp \matrix{1\cr 0}\rp \ ,
\quad \psi_2 =  e^{ky-\varphi(y)} \lp \matrix{0 \cr 1} \rp \ .
\end{equation}
%where $c_1$ and $c_2$ are constants. 
Note that since $k$ is real
and $\varphi(y)$ is periodic in $y$, the zero-energy solutions (\ref{indep-solns})
are compatible with the boundary condition~(\ref{def-BC}) 
{\sl if and only if} $\delta=0$ and $k=0$.
For the latter case the exact zero-energy eigenstates of
$\tilde\H_D$ can be written as
\be
\psi_1^{(0)}\!=\! {e^{\varphi(y)}\over \sqrt{L g_0^{(1)}}}
\lp \matrix{1\cr 0}\rp \ ,
\quad
\psi_2^{(0)}\!=\! {e^{-\varphi(y)}\over \sqrt{L g_0^{(2)}}}
\lp \matrix{0 \cr 1} \rp \ ,
\label{eq:zero-energy-solutions}
\ee
where normalization requires 
\be \label{def-g-general}
Lg_0^{(1)} = \int_0^L \!\! dy\, e^{2\varphi(y)} \ , \quad
Lg_0^{(2)} = \int_0^L \!\! dy\, e^{-2\varphi(y)} \ .
\ee
In the case of a uniform perpendicular field
\be \label{def-varphi}
\varphi
%% = {eR\over \hbar c}\,\int_0^\theta \! d\theta' \, A(\theta')
= 2\phi\, \sin \theta \ ,
\quad \theta =y/R \ ,
\ee
the normalization factors are given by the modified Bessel function:
%%$I_0(z)=\frac1{2\pi}\oint e^{z\sin\theta}d\theta$.
\be \label{g0Bessel}
g_0^{(1)}=g_0^{(2)}=\frac1{2\pi}\oint e^{4\phi \sin\theta}d\theta  = I_0(4\phi) \ .
\ee
The states (\ref{eq:zero-energy-solutions})
are degenerate at any field strength.

Let us stress here that the zero-energy eigenstates (\ref{eq:zero-energy-solutions}) 
of $\tilde\H_D$ exist \cite{Dubrovin} 
at $\delta=0, \ k=0$ for a generic magnetic field 
perpendicular to the NT and not varying along the tube.
An example is a field of a current flowing along the wire parallel to the 
nanotube axis. Moreover, the above arguments still apply 
if the NT does not have a circular cross section, 
as long as the minigap due to the
curvature\cite{Hamada92PRL,Kane97PRL} of the graphene sheet is not open.
In what follows we confine ourselves to
the case of a cylindrical NT in a homogeneous perpendicular magnetic 
field for simplicity, 
bearing in mind the generalizations mentioned above.
%%Eqs.~(\ref{eq:zero-energy-solutions},\ref{def-varphi}) in a more
%%general case.

\begin{figure}[t]
\includegraphics[width=3.5in,height=3in]{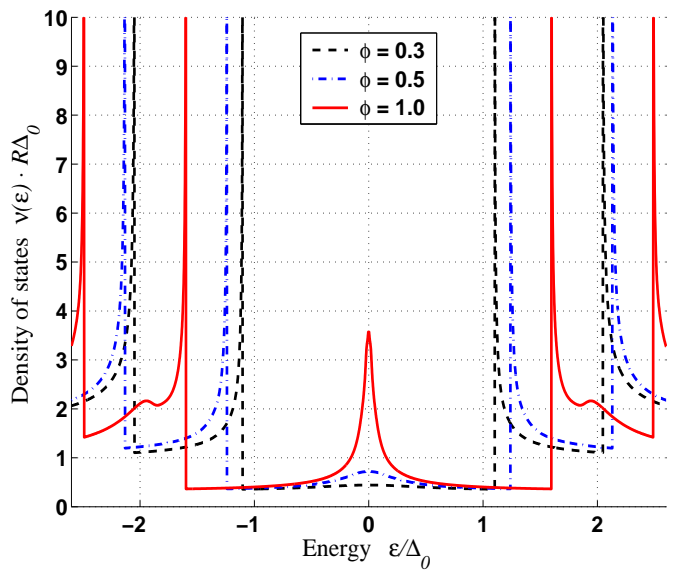}
\caption[]{\label{fig:dos}
Density of states $\nu=dN/d\epsilon$ in metallic NT's per one fermion flavor
as a function of energy in the units of $\Delta_0 = \hbar v/R$.
The peak value $\nu(0)$ at the band center is given by Eq.~(\ref{eq:nu}).
}
\end{figure}

Using the states (\ref{eq:zero-energy-solutions}) one can
study how the linear dispersion relation changes near the band center
$\epsilon=0$. For that
we project the Hamiltonian (\ref{H-dirac}) onto the basis
\be\label{eq:Psi12basis}
\psi_{1,2}(x,\theta) = e^{ikx} \psi_{1,2}^{(0)}(\theta) \ .
\ee
The projected Hamiltonian 
\be \label{H-D-projected}
{\H_D}|_{\Psi_{1,2}} = {\hbar k\bar v} \,\sigma_2  \ , 
\quad \bar v = {v \over I_0(4\phi)},
\ee
yields the dispersion relation
\be \label{g-metallic}
\epsilon(k) = \pm \hbar k \lp {v\over I_0(4\phi)} \rp \ .
\ee
This describes a {\it reduction} of the
Fermi velocity $\hbar^{-1}d\epsilon/dk$ near $\epsilon=0$ by a factor $I_0(4\phi)$.
Since $I_0(4\phi)>1$, the density of states at the band center,
\be\label{eq:nu}
\nu=dN/d\epsilon={\textstyle\frac4{\pi\hbar v}}\,I_0(4\phi),
%%/(\pi\hbar v)
\ee
is {\it enhanced}.
(The factor of $4$ accounts for the spin
and valley degeneracy neglecting the Zeeman splitting; see
Sec.~\ref{sec:gap-suppression} below.)
Due to the exponential behavior of the Bessel
function in Eq.~(\ref{g-metallic}) at large $\phi$, this enhancement becomes
dramatic at high fields (Fig.~\ref{fig:dos}).
The reduction of the Fermi velocity and the corresponding density-of-states
enhancement is a general consequence of a supersymmetry.
Indeed, in the generic field case described above
the velocity is reduced by a factor
$\sqrt{g_0^{(1)}g_{0}^{(2)}} \geq 1$.
Both $g_{0}^{(1)}, \,g_{0}^{(2)}\geq 1$  [see Eq.~(\ref{def-g-general})]
since the average of an exponential is greater than or equal to
an exponential of the average.

%%%%%%%%%%%%%%%%%%%%%%%%%%%%%%%%%%%%%%%%%%%%%%%%%%%%%%%%%%%%%%%%
%%%%%%%%%%%%%%%%%%%%%%%%%%%%%%%%%%%%%%%%%%%%%%%%%%%%%%%%%%%%%%%%

In the single-particle approximation, the tunneling density of states
coincides with the thermodynamic density of states (\ref{eq:nu}).
The peak in the tunneling density of states at the band center
is a distinct manifestation of supersymmetry. Recently, a scanning tunneling
probe has been used \cite{Ouyang01} to study curvature-induced minigaps in
nominally metallic tubes placed on a metallic substrate.
In this setup, the electron interactions that could have modified the
single-particle behavior are screened by the substrate, and the measured
density of states is unaffected by Luttinger liquid effects.
In a similar system in a high transverse field, an enhancement
of tunneling at the band center, Eq.~(\ref{eq:nu}), and a peak in the
density of states (Fig.~\ref{fig:dos}) are the experimental signatures
to look for.

%%%%%%%%%%%%%%%%%%%%%%%%%%%%%%%%%%%%%%%%%%%%%%%%%%%%%%%%%%%%%%%%
%%%%%%%%%%%%%%%%%%%%%%%%%%%%%%%%%%%%%%%%%%%%%%%%%%%%%%%%%%%%%%%%
\section{Gap suppression}
%%%%%%%%%%%%%%%%%%%%%%%%%%%%%%%%%%%%%%%%%%%%%%%%%%%%%%%%%%%%%%%%
%%%%%%%%%%%%%%%%%%%%%%%%%%%%%%%%%%%%%%%%%%%%%%%%%%%%%%%%%%%%%%%%
\label{sec:gap-suppression}

%\nin
In the present section we consider nanotubes 
that have a gap $\Delta$ at the band center.
We characterize them by the parameter $|\delta| = \Delta/\Delta_0$
that enters the boundary condition (\ref{def-BC}).
There are two kinds of such tubes: 
semiconducting NT's with $\delta=\pm1/3$ and nominally metallic
NT's in which a minigap appears due
to curvature or external field, 
yielding a small $|\delta| \ll 1$. The Dirac problem
(\ref{H-dirac}) and (\ref{def-BC}) is not supersymmetric for
$\delta\ne0$. However, since the supersymmetry is an exact property at
$\delta=0$, one can still expect it
to manifest itself in a problem with a relatively small $\delta$.
Below we show that a gap at the band center is {\it suppressed} in the presence of
an external transverse field:
\be\label{def-g}
\Delta(\phi)=\frac{\Delta}{g_\delta(\phi)}\ , \quad g_\delta(\phi) > 1  \ ,
\ee
where the gap suppression factor $g_\delta(\phi)$ diverges as
$\phi\rightarrow \infty$.
This means that supersymmetry is restored in the limit of a strong
field. The suppression of the gap in semiconducting NT's is illustrated in
Fig.~\ref{fig:dispersion} (lower panel).

We shall first consider a simpler case of a nominally metallic
tube with $\delta\ll1$. The gap in this case can be found using perturbation
theory in $\delta$. For that, we perform a gauge transformation
$\psi(\theta)\to e^{i\delta\theta}\psi'(\theta)$ and for $\psi'(\theta)$
obtain a problem with the {\sl periodic} boundary condition. The new
Hamiltonian differs from Eq.~(\ref{H-dirac}) by a term linear in $\delta$:
\be
{\cal H}'={\cal H}_D - \delta \Delta_0\sigma_1 \ .
\ee
It is convenient to rewrite this Hamiltonian as
\be\label{eq:H'(k)}
{\cal H}'={\cal H}_D^{(0)}+\Delta_0(kR\,\sigma_2 - \delta\sigma_1)
\ ,
\ee
where ${\cal H}_D^{(0)}$ is the Hamiltonian (\ref{H-dirac})
with $k=0$. We note that ${\cal H}_D^{(0)}$ is a supersymmetric Hamiltonian
with eigenstates
(\ref{eq:zero-energy-solutions}). The spectrum of the Hamiltonian
${\cal H}'$ at small $kR$ and $\delta$ can be found by projecting
the second term of Eq.~(\ref{eq:H'(k)}) on the basis (\ref{eq:Psi12basis})
of plane-wave states constructed out of
Eq.~(\ref{eq:zero-energy-solutions}). This yields the dispersion relation
\be
\epsilon(k)=\pm \frac{\Delta_0}{I_0(4\phi)}\left[
(kR)^2+\delta^2\right]^{1/2} \ .
\ee
Thus we find that in this case the gap is suppressed
by the same factor (\ref{g0Bessel}),
\be\label{eq:g0Bessel}
g_0(\phi)=I_0(4\phi) > 1,
\ee
as the Fermi velocity in metallic NT's. The gap suppression
is described by Eq.~(\ref{eq:g0Bessel}) in the limit of small $\delta$
for any magnetic field $\phi$.

One can also study the gap suppression analytically for generic
$\delta$ using perturbation
theory in the field $\phi$. The energy $\epsilon(k=0)$
that gives the gap is defined by the condition (\ref{eq:trace-criterion})
for the transfer matrix.
We calculate the trace of the transfer matrix (\ref{def-S}) at $k=0$
by expanding it perturbatively in $\phi\ll 1$:
\be
{\rm tr\,}\, S_{\theta=2\pi} = 2\cos 2\pi\bar\epsilon  + 8\lambda_1\, \phi^2 +
32\lambda_2\, \phi^4 + {\cal O}(\phi^6),
\ee
where $\tilde\epsilon = \epsilon(0)/\Delta_0$ and the coefficients
$\lambda_{1,2}$ are given by
\bea
\label{def-lambda-1}
\!\!\!\lambda_1 \! & = &\! - \frac{2\pi\tilde\epsilon \,
\sin 2\pi\tilde\epsilon}{1-4\tilde\epsilon^2} \ ,  \\
\!\!\!\lambda_2 \! & = &\! -\pi\tilde\epsilon \;
\frac{
2\pi\tilde\epsilon \left(1-4\tilde\epsilon^2\right)
\!\cos 2\pi\tilde\epsilon
+ \lp \! \frac12+6\tilde\epsilon^2 \rp \!\sin 2\pi\tilde\epsilon}
{\lp 1-4\tilde\epsilon^2 \rp^3} \ .
\label{def-lambda-2}
\eea
The condition (\ref{eq:trace-criterion}) on the energy along with
the definition of the suppression factor (\ref{def-g}) gives
\bea
\label{def-g-4}
g_{\delta} &=& 1 + \alpha_{\delta}\phi^2
+ \beta_{\delta}\phi^4  + {\cal O}(\phi^6)\ ,  \\
\label{def-alpha}
\alpha_{\delta} &=& {4 \over 1 - 4\delta^2} \ , \quad
\beta_{\delta} = 4\, \frac{1 - 20\delta^2}{\left(1-4\delta^2 \right)^2} \ .
\eea
Substituting, in Eq.~(\ref{def-alpha}),  $\delta=0$ and $\delta=1/3$ we obtain
\bea
\label{g-met-approx}
\! \! g_{0}(\phi) &=& 1 + 4\phi^2 + 4\phi^4 + {\cal O}(\phi^6)\quad
{\rm (metallic)} \ , \\
\label{g-sc}
\! \! \!\! \! g_{1/3}(\phi) &=& 1 + \frac{36}5 \phi^2  - \frac{396}{25} \phi^4
+ {\cal O}(\phi^6) \quad \!
{\rm (semiconducting)} .
\eea
The expression (\ref{g-met-approx}) coincides with the Taylor expansion of
$I_0(4\phi)$.

\begin{figure}[t]
\includegraphics[width=3.5in,height=3in]{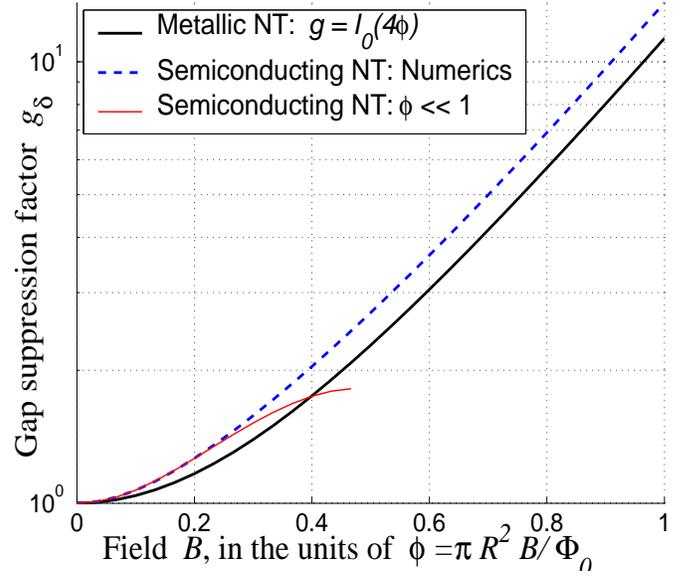}
\vspace{0.5cm}
\caption[]{\label{fig:suppression-factor}
Gap suppression factors (\ref{def-g}) for nominally metallic
NT's, $\delta=0$ (solid line), and semiconducting NT's, $\delta=1/3$
(dashed line), as a function of field $\phi$.
The fine solid line is the result of the expansion
(\ref{g-sc}) in powers of $\phi$. (Note the logarithmic scale for $g_{\delta}$.)
}
\end{figure}

These analytical results can be compared with the
gap suppression factors obtained numerically
(Fig.~\ref{fig:suppression-factor}).
For nominally metallic NT's
with small minigap we find that at $\delta\ll 1$ the value
$g_{\delta\to 0}$ is accurately given
by Eq.~(\ref{eq:g0Bessel}). The analytical expression (\ref{eq:g0Bessel})
coincides with the numerics in the entire field range.
In the semiconducting case of $\delta=1/3$ the expansion (\ref{g-sc})
works reasonably well at $\phi \le 1/4$. At larger fields $\phi>1/4$
the gap is suppressed exponentially,
$g_{1/3}(\phi)\propto e^{4\phi}$ (see Fig.~\ref{fig:suppression-factor}).

%%%%%%%%%%%%%%%%%%%%%%%%%%%%%%%%%%%%%%%%%%%%%%%%%%%%%%%%%%%%%%%%
%%%%%%%%%%%%%%%%%%%%%%%%%%%%%%%%%%%%%%%%%%%%%%%%%%%%%%%%%%%%%%%%

Let us discuss the possibilities to observe the suppression of the gap.
A competing effect due to the magnetic field that leads to a gap suppression is
the Zeeman spin coupling ${\cal H}_Z=-\mu \sigma B$
with $\mu=e\hbar/2m c$.
The gap suppression at weak fields, 
$\phi\ll1$, with the Zeeman effect included, is described by
\be\label{eq:delta(B)}
\Delta(\phi)=\Delta-\alpha_\delta\phi^2\Delta-\mu B,
\ee
with $\alpha_\delta$ given by Eq.(\ref{def-alpha}). The Zeeman effect,
linear in $B$, dominates at weak fields. However, the orbital effect
$\alpha_\delta\phi^2\Delta$, quadratic in $B$,
overcomes the Zeeman effect at relatively moderate fields
$\phi\ll1$. 

For semiconducting NT's the gap $\Delta = \hbar v/3R$, and Eq.~(\ref{def-alpha})
gives $\alpha_{1/3}=36/5$. In this case, the inequality
$\alpha_{1/3}\phi^2\Delta > \mu B$
yields
$\phi > 1.13\, a_B/R$
with $a_B = \hbar^2/m e^2$ the Bohr radius.
Using carbon parameters we estimate that
the magnetic field has to exceed
\be \label{def-B*}
B_0 = 78\,(R\,{\rm [nm]})^{-3} \,\,\,\, {\rm T }
\ee
which can be low enough for tubes of large radius.

The energy gap can be studied experimentally by measuring transport
in the thermally activated regime. The thermal activation energy
will depend on the magnetic field according to Eq.~(\ref{eq:delta(B)}).
The resistance
\be \label{activ-cond}
R(T)\propto \exp[\Delta(\phi)/k_{\rm B} T]
\ee
will be sensitive to magnetic field because the variation of the gap
can exceed $k_{\rm B} T$
even at the fields much smaller than Eq.~(\ref{def-B*}).
For example, consider a NT of radius $R=1\,{\rm nm}$, in which case
$\Delta=\hbar v/3R=0.178\,\,\,{\rm eV}$.
For the magnetic field $B=B_0$ from Eq.~(\ref{def-B*}) we have
$\phi=0.060$ and
$\alpha_{1/3} \phi^2 = 0.026$. In this case the gap shift
$\alpha_{1/3}\phi^2\Delta=4.6\,{\rm meV}$ is larger than $k_{\rm B} T$
at $T<53\,{\rm K}$.

%%%%%%%%%%%%%%%%%%%%%%%%%%%%%%%%%%%%%%%%%%%%%%%%%%%%%%%%%%%%%%%%
%%%%%%%%%%%%%%%%%%%%%%%%%%%%%%%%%%%%%%%%%%%%%%%%%%%%%%%%%%%%%%%%
\section{Electron spectrum in extremely large fields}
\label{sec:snake}

%\nin 
Below we consider the qualitative features
of the energy bands in metallic NT's in the limit
of a large uniform external field, corresponding to $\phi\geq 1$. 
The behavior of the band dispersion at
$|k|R\ll \phi$ can be attributed to the Landau levels
of the problem (\ref{H-dirac}). Let us consider a square
of the Hamiltonian (\ref{H-dirac})
and the eigenvalue problem ${\cal H}_D^2\psi=\epsilon^2\psi$.
It reads
\bea \label{def-HD2}
\H_D^2 \psi = \Delta_0^2 [ - \partial_\theta^2 + U_k(\theta) ] \psi
= \epsilon^2(k) \, \psi \ , \\
\label{def-U}
U_k(\theta) = (2\phi\cos \theta - kR)^2 - 2\phi\sigma_3 \sin \theta \ ,
\eea
with $\psi(\theta)$ obeying the boundary conditions (\ref{def-BC}).
Note that ${\cal H}_D^2$ is a diagonal $2\times2$ matrix
in the space of spinors $\psi$.
In what follows we take $\sigma_3 = +1$.

When $\phi\gg 1$, the kinetic energy of Eq.~(\ref{def-HD2}) is frozen since 
$U_k(\theta)\propto \phi^2$ is much greater than $\partial_\theta^2$.
In this limit, the Hamiltonian is dominated
by the potential energy term $U_k(\theta)$ and the low-energy states
are localized near the minima of $U_k(\theta)$.
At $\phi \gg 1$ the potential $U_k(\theta)$ has two slightly
asymmetric minima near $\theta_{\pm}=\pm \pi/2$, where
it can be approximated by a harmonic potential
$U_k(\theta)\approx \mp 2\phi + 4\phi^2 (\theta-\theta_{\pm})^2$.
The size of electronic wave function in the circumferential direction
is
\be \label{mag-len}
w \simeq l_B  = R/\sqrt{2\phi} \ll R \ ,
\ee
where $l_B = (\hbar c / eB)^{1/2}$ is the magnetic length.
Thus at large field the electrons are localized near
the extrema of the magnetic field $\theta_{\pm}$. In these regions
the field
is approximately constant: $|B_{\perp}(\theta)|\approx B$.

The Landau level spectrum $\epsilon_n^2(k)$ of Eq.~(\ref{def-HD2})
obtained within the harmonic approximation
yields approximately $k$-independent levels for $\H_D$: 
\be \label{def-landau-levels}
\epsilon_n(kR\ll\phi) = \pm 2 \, \Delta_0 \, \sqrt{n |\phi|} \ ,
\quad n=0,1,2,...~.
\ee
The lowest-energy level of Eq.~(\ref{def-HD2}) is
$\epsilon=0$ at $k=0$. This value, because of the supersymmetry,
is not approximate but exact.

The behavior at large momenta
$|k|R\gg \phi$ can be understood semiclassically in terms of
the so-called {\sl snake states}.~\cite{snake} Snake states
correspond to a classical particle moving along zero-field lines.
This motion is stable for a particle traveling in one direction 
and is unstable for it traveling in the opposite one.
The snake states are located at $\theta=0$ and $\theta=\pi$, where 
the field $B(\theta)=B\sin\theta$ vanishes.
This is consistent with the high-field limit of
the problem (\ref{def-HD2}), since at $|k|R\gg \phi$
the minimum of $U_k(\theta)$ is
$\theta_0 \approx 0$ for $kB>0$ and $\theta_0 \approx \pi$ for $kB<0$.
The dispersion relation for such states is
\be
\epsilon(k) = \pm \Delta_0 \sqrt{U_k(\theta_0)} \approx \pm \hbar v
(|k| - 2|\phi|/R) \ .
\ee
This linear dispersion relation with an offset $2|\phi| \, \hbar v/R$
holds even for small fields as long as
$|k|R \gg {\rm max\,}\lf\phi,\,1/\phi\rf$ 
(see Fig.~\ref{fig:dispersion}, large $kR$).

Since $\theta_0$ is different for positive and negative $k$,
the left- and right-moving snake states are spatially separated.
For $\phi>0$, for instance,
the left- (right-) moving snake states are localized near
$\theta=\pi$ ($\theta=0$).
The characteristic width of the snake-state wave function is
\be \label{def-w-snake}
w_{\rm snake} = l_B^{1/2} (R/k)^{1/4} \propto |\nabla B|^{-1/4} \ .
\ee
The width $w_{\rm snake} \ll R$ when $kR \gg 1/|\phi|$.
In the Dirac problem the wave function width
(\ref{def-w-snake}) is different from that for the Schr\"odinger problem
discussed in Ref.~\cite{snake}, where
$w_{\rm snake} \propto |\nabla B|^{-1/2}$.

%%%%%%%%%%%%%%%%%%%%%%%%%%%%%%%%%%%%%%%%%%%%%%%%%%%%%%%%%%%%%%%%
%%%%%%%%%%%%%%%%%%%%%%%%%%%%%%%%%%%%%%%%%%%%%%%%%%%%%%%%%%%%%%%%
\section{Beyond the Dirac equation: supersymmetry breaking}
\label{sec:beyond-dirac}

%\nin
The two effects considered in Secs. \ref{sec:dirac} 
and \ref{sec:gap-suppression}---the density of states enhancement 
at $\epsilon=0$ in metallic NT's
and the suppression of the energy gap in semiconducting NT's---are 
manifestations of the supersymmetry
of the low-energy Dirac Hamiltonian (\ref{H-dirac}).
However, supersymmetry is not present
in the original tight-binding problem (\ref{eq:2Dgraphite}).
Below we show that the terms
correcting Eq.~(\ref{H-dirac}) in the next order of the gradient expansion
violate supersymmetry. Thus the supersymmetry in nanotubes is not exact
but approximate: the nonsupersymmetric effects are small in  $a_{\rm cc}/R$.

To obtain the nonsupersymmetric terms of the effective Hamiltonian
we consider the low-energy subspace of states
with $|\epsilon| \ll t$ near the Dirac points $K$ and $K'$.
The basis states at $\epsilon=0$ are the functions $u(r)$, $v(r)$ and
$\bar u(r)$, $\bar v(r)$ defined in Sec.~\ref{sec:carbon-basic}
(see Fig.~\ref{fig:uv-states}). The wave function near the point $K$ ($K'$)
can be represented as a linear superposition (\ref{eq:psi12})
of $u(r)$ and $v(r)$ [respectively, $\bar u(r)$ and $\bar v(r)$]
multiplied by the smooth envelope functions $\psi_{1,2}(r)$.
We perform a gradient expansion of
the slowly varying envelope functions using
Eq.~(\ref{eq:p1p2z}).
In the lowest nonvanishing order in $a_{\rm cc}\partial\psi_{1,2}$
we retain the Hamiltonian (\ref{H-dirac}) with $v=3ta_{\rm cc}/2\hbar$.

The terms of second order in the gradients
$a_{\rm cc}^2\partial_i\partial_j\psi_{1,2}$ give the required
correction $\H_{\rm tw}$ to the Dirac Hamiltonian $\H_D$
called the carbon {\it trigonal warping} interaction. In this case,
since we are interested in the problem in an external field,
the gradient expansion of Eq.~(\ref{eq:p1p2z})
should be accompanied by an expansion of the phase factors
(\ref{eq:gamma-rr'}). After this expansion is carried out we choose the
tube axis orientation with respect to the carbon lattice by specifying
the chiral angle $\Theta$.
The full Hamiltonian $\H_{\rm tot}$ obtained in such a way for NT's
has the form
\be \label{def-H-tot}
\H_{\rm tot}= e^{i(\Theta/2) \sigma_3} \H_D e^{-i(\Theta/2) \sigma_3}
 +e^{-i\Theta \sigma_3} \H_{\rm tw} e^{i\Theta \sigma_3} \ ,
\ee
where $\H_D$ is the Dirac Hamiltonian (\ref{H-dirac})
and the trigonal warping interaction $\H_{\rm tw}$ is given by
\be \label{def-H-a}
\H_{\rm tw} = -{a_{\rm cc}\over 4R} \, \Delta_0
\lf
\lp \kappa^2 + \partial_{\theta}^2 \rp \sigma_1
\!+\! i\lp\!
2\kappa \, \partial_{\theta}
\! +\! {d \kappa \over d \theta}
\!\rp \sigma_2
\rf, 
\ee
where
\be \quad
\kappa(\theta) \equiv kR  - {eR\over \hbar c}A(\theta) \ .  \quad
\ee
\nin
The term $\H_{\rm tw}$ breaks the supersymmetry of the Hamiltonian $\H_{\rm tot}$.
Thus we expect the zero-energy state to disappear.
Note that $\H_{\rm tw}$ also breaks the rotational symmetry
of $\H_{\rm tot}$ since $\Theta$ cannot be removed from
$\H_{\rm tot}$ via a unitary transformation.
Thus the behavior of the energy gap in a transverse field
will in general depend on $\Theta$. It can be verified that,
in the absence of external fields, the energy spectrum
of $\H_{\rm tot}$ is periodic
in $\Theta$ with period $\pi/3$,
which is a manifestation of the $60^\circ$ rotation symmetry
in the honeycomb lattice.

It is explicit in Eq.~(\ref{def-H-a}) that the effects of $\H_{\rm tw}$
are the order of $a_{\rm cc}/R$ corrections to $\H_D$.
Such effects are negligible
for semiconducting NT's because of a large gap $\Delta_0/3$.
In metallic NT's, however, the Hamiltonian $\H_{\rm tw}$ plays an important
role.
In particular,
the system described by the Hamiltonian (\ref{def-H-tot})
develops a minigap
\be \label{def-delta-tw}
\Delta_{\rm tw}={2a_{\rm cc} \over R}\Delta_0\,
{\phi^2 |\cos 3\Theta| \over g_0(\phi)}
\ee
due to the magnetic field $\phi$ [defined in Eq.~(\ref{def-A})].
This result can be obtained by projecting the perturbation
$\H_{\rm tw}$ taken at $k=0$ onto the supersymmetric
basis (\ref{eq:zero-energy-solutions}):
\be
\H_{\rm tw}|_{k=0} = - {2a_{\rm cc} \over R}\Delta_0\,
{\phi^2 \over g_0(\phi)} \ \sigma_1 \ .
\ee
Note that the minigap (\ref{def-delta-tw}) depends explicitly on
the chiral angle $\Theta$ as $|\cos 3\Theta|$.
Thus, for a given NT radius, $\Delta_{\rm tw}$ reaches
its maximum in zigzag NT's ($\Theta=0$) and vanishes in armchair NT's
($\Theta=\pi/2$).
The gap (\ref{def-delta-tw}) is a manifestation of
the broken supersymmetry.
Minigaps of purely magnetic origin have been
reported in Ref.~\cite{Ajiki96JPSJ} for zigzag and armchair NT's;
however, the gap dependence on the chiral angle $\Theta$
was not discussed.

When the magnetic field is large, $\phi \sim 1$,
the minigap $\Delta_{\rm tw}$ is comparable
to the curvature-induced minigap
\be \label{def-curv-gap}
\Delta_{c}={a_{\rm cc}\over 16 R}\, \Delta_0  |\cos 3\Theta|,
\ee
which is present in {\sl zero} field.~\cite{Hamada92PRL,Kane97PRL} 
In this situation, the
two mechanisms for minigaps may compete and should be considered
simultaneously. Instead of imposing the quasiperiodic boundary
condition (\ref{def-BC}), we take the curvature effect into
account in an alternative way~\cite{Kane97PRL} by introducing a
pseudovector potential $\vec A^{(c)}$, \be \label{def-A-c}
A^{(c)}_x+iA^{(c)}_y = i\, {\Phi_0 \over 2\pi} \cdot {a_{\rm
cc}\over 16R^2}\, e^{3i\Theta} \ , \ee which should be added to
the magnetic vector potential ${\bf A}(\r)$. Surprisingly, the two
gap opening mechanisms, Eqs.~(\ref{def-H-a}) and (\ref{def-A-c}),
interfere {\it destructively} at both the $K$ and $K'$ points and
produce the gap \be \Delta_\phi= \left| {\Delta_c \over g_0(\phi)}
- \Delta_{\rm tw} \right| \label{Delta-phi} \ee
in a moderate transverse magnetic field ($\phi \lesssim 1$). In
particular, $\Delta_\phi$ {\it vanishes} at
$\phi=(4\sqrt{2})^{-1}$ due to the destructive interference.

%%%%%%%%%%%%%%%%%%%%%%%%%%%%%%%%%%%%%%%%%%%%%%%%%%%%%%%%%%%%%%%%
%%%%%%%%%%%%%%%%%%%%%%%%%%%%%%%%%%%%%%%%%%%%%%%%%%%%%%%%%%%%%%%%
\section{Electron Interactions {\it vs.} SUSY}
\label{sec:inter}

%\nin 
Electron interaction effects on NT's were addressed in various 
theoretical~\cite{CNT-Mott,Levitov'01,CNT-LL,Talyanskii'01,Lee03PRL} 
and experimental~\cite{CNT-LL-exp} studies.
Below we consider the effects of the repulsive interaction between electrons
on the NT spectrum in the presence of a perpendicular magnetic field.
As we have seen above in Secs.~\ref{sec:dirac} and \ref{sec:gap-suppression}, 
supersymmetry of the single-particle problem makes 
nanotubes ``more metallic'' in a strong field by enhancing the density of 
states and suppressing the excitation gap at zero energy in metallic and 
semiconducting tubes correspondingly.
It is also known that the repulsive interaction between NT electrons
opens a small
gap in an otherwise metallic tube \cite{CNT-Mott} as well as 
enhances the excitation gap in a semimetallic tube,~\cite{Levitov'01}
making the tubes ``less metallic.''
Below we study the competition between supersymmetry and 
repulsive electron interactions, and find that
strong interactions drastically reduce the effect of supersymmetry.
This happens because supersymmetry enhances electron interactions near half-filling
as one would expect from the increase in the density of states (\ref{eq:nu}).

In what follows we consider the case of a very strong magnetic
field $\phi > 1$. The latter condition corresponds to an 
exponentially large effect of the field on the single-electron NT spectrum.
For that reason we will neglect the Zeeman effect, which is
linear in field.

We consider the interacting problem whose Hamiltonian 
in the forward scattering approximation
\cite{CNT-LL} reads
\be \label{Htot}
\H_{\rm tot} = \H_0 + \H_{\rm int} \ , 
\ee
where 
\be \label{H0}
\H_0 = \hbar v \int\! d\vec{r}\, \sum_{\alpha=1}^4 \Psi_{\alpha}^{\dagger}
\left\{ \lp i\partial_y - {\delta\over R} \rp\sigma_1 
- (i\partial_x + \varphi'_y )\sigma_2 \right\} \Psi_{\alpha} 
\ee
and 
\be \label{Hint}
\H_{\rm int} = {1\over 2}\, \sum_k   
\rho_{-\k} V(\k) \rho_{\k} \ .
\ee
Here the Hamiltonian (\ref{H0}) describes four noninteracting fermion flavors
($4 = 2_{\rm spin}\times 2_{\rm valley}$)
in a nanotube with the bare gap $\Delta = \hbar v \delta/R$, subject to a
perpendicular magnetic field, $\varphi(y) = 2\phi \sin (y/R)$, 
$\varphi'_y \equiv d\varphi/ dy$, where 
the dimensionless field strength $\phi$ is defined in Eq.~(\ref{def-A}).
The Dirac spinors $\Psi_{\alpha}$ are operators in 
the second-quantized representation.
The Coulomb interaction between electrons is described by 
the Hamiltonian (\ref{Hint}), where the total density in the 
forward scattering approximation reads 
\be
\rho(\r) = \sum_{\alpha=1}^4 \Psi_{\alpha}^{\dagger}\Psi_{\alpha} \ ,
\ee
with the $2K_0$ harmonics 
[$K_0$ defined in Eq.~(\ref{K0})] neglected.
The electron-electron interaction potential 
in the presence of a substrate with a dielectric constant $\varepsilon$ is 
\be \label{V-2d}
V(x,y) = \textstyle{\frac{2}{\varepsilon+1} V_0(x,y)} \ ,
\ee
where
\be \label{V0-2d}
V_0(x,y)={e^2\over \sqrt{h^2 + x^2}}\ ,  \quad h=2R \sin (y/2R) \ .
\ee
The problem (\ref{Htot}) is SU(4) invariant with respect to 
rotations in the space of the four fermion flavors $\Psi_{\alpha}$.

Below we focus on the low-energy properties of the problem (\ref{Htot}).
This allows us to utilize the projection on the supersymmetric basis 
(\ref{eq:zero-energy-solutions}):
\be \label{projection}
\Psi_{\alpha}(\r) = \chi_{1 \, \alpha}(x) \psi_{1}^{(0)}(y) 
+ \chi_{2\, \alpha}(x) \psi_{2}^{(0)}(y) \ , 
\ee
where the factorization of motion along $x$ and $y$
holds due to the assumption that the magnetic field does not vary along the tube. 
Using Eq.~(\ref{projection}) 
we will reduce the problem (\ref{Htot}) to the one-dimensional one, 
bosonize it, and estimate the effect of interactions on the plasmon velocity
and on the semimetallic gap. 

Let us perform a projection of the problem (\ref{Htot}) onto the 
basis (\ref{projection}). This can be done by integrating out the circumferential
degree of freedom using the following separation of scales. The
effects of magnetic field occur on the short scale of the tube radius $R$,
for which the relevant energy scale is $\sim \hbar v /R$.
The effect of the Coulomb interaction between electrons accumulates over 
a length scale that is much greater than $R$, as described below.
Therefore the effective 1D description of the interacting NT electrons
can be obtained by first integrating out the circumferential
coordinate $y$ in Eq.~(\ref{Htot}) and then taking into account the Coulomb effects.
Thus we obtain the effective one-dimensional Hamiltonian  
\bea \nonumber
\H_{\rm eff} &= & {\hbar \bar v}
\int\! dx \sum_{\alpha} \chi_{\alpha}^{\dagger}
\lp -i\partial_x\sigma_2 -{\delta\over R}\,\sigma_1\rp  \chi_{\alpha} \\ 
&&+ {1\over 2}\, \sum_k \tilde\rho_{-k} \tilde V(k) \tilde \rho_{k} \ ,
\label{Heff}
\eea
with the bare velocity reduced due to supersymmetry,
\be
\quad \bar v = {v\over g_0(\phi)} \ , 
\ee
similar to Eq.~(\ref{H-D-projected}).
Here $g_0$ is given by Eq.~(\ref{eq:g0Bessel}), 
\be \label{chi}
\chi_{\alpha} = \lp \matrix{\chi_{1\, \alpha} \cr \chi_{2\, \alpha} }\rp \ , 
\ee
the one-dimensional electron density (calculated from half-filling)
\be
\tilde \rho = \sum_{\alpha=1}^4 \chi_{\alpha}^{\dagger}\chi_{\alpha} \ ,
\ee
and the 1D interaction potential at $kR\ll 1$
\be \label{t-V}
\tilde V(k) \simeq  \frac{2e^2}{\varepsilon+1} \ln \lb 1 + (kR)^{-2}\rb \ . 
\ee
In writing Eq.~(\ref{Heff}) we dropped the terms of the order 
$[e^2/ (\varepsilon+1)] \ln \phi$ that are small compared to $\tilde V(k)$
at $kR\ll 1$. 
These terms appear since at $\phi>1$ the states $\chi_{1\, \alpha}$ and 
$\chi_{2\, \alpha}$ are localized on the opposite sides of the tube. 
Therefore strictly speaking 
the interaction between the same components of the spinor (\ref{chi})
is cut off on the scale of magnetic length (\ref{mag-len}) rather than of the
NT radius.

With the difference between the short-distance cutoffs 
in the potential (\ref{t-V}) neglected, 
the effective 1D Hamiltonian (\ref{Heff}) remains SU(4) invariant. 
It can be bosonized in the standard way,~\cite{CNT-LL}
$\chi_{\alpha} \propto e^{i\Phi_{\alpha}}$. 
This procedure 
immediately yields the renormalized plasmon velocity for a metallic nanotube,
\be \label{t-v}
\tilde v = K^{1/2}(\phi) {v \over g_0(\phi)} \ , 
\ee
where the charge stiffness (or the dimensionless interaction strength)
\be \label{K}
K_q(\phi) = 1 + {4 g_0(\phi)\over \pi \hbar v}\, \tilde V(q) 
\ee
is {\sl enhanced} by the magnetic field.
(By the tilde we denote the physical quantities in the presence of 1D interactions.)
Thus the plasmon velocity suppression factor 
$\tilde{g}_0(\phi)\equiv[\tilde{v}(\phi)/ \tilde{v}(0)]^{-1}$
due to the magnetic field is given by
\be \label{new-suppression-factor}
\tilde{g}_0(\phi)=\left[ {K(0) \over K(\phi)} \right]^{1/2} g_0(\phi) \ .
\ee
It is {\sl reduced} compared to the noninteracting value $g_0(\phi)$
because of the enhancement of the interaction strength due to
the perpendicular magnetic field. For a large interaction strength $K\gg 1$,
the effect of electron interactions on the supersymmetry is dramatic:
\be
\tilde{g}_0(\phi) \simeq \lb g_0(\phi)\rb^{1/2} \propto e^{2\phi} \ ,
\ee
effectively reducing the field strength $\phi > 1$ by a factor of 2 in the 
exponential.

Consider now the semimetallic gap in the presence of a magnetic field.
In bosonized language, this gap is estimated as the energy of 
a composite soliton of the charge and flavor modes,~\cite{Talyanskii'01,Levitov'01}
with its energy dominated by that of the charge mode at $K\gg 1$. 
The essential feature for the present analysis is that 
the effect of the magnetic field on the Gaussian part of $\H_{\rm eff}$ 
{\sl factorizes}, renormalizing the velocity $\bar v$,
with the backscattering term 
$\delta/R \equiv \Delta(\phi) g_0(\phi) /\hbar v$
inside the integral in Eq.~(\ref{Heff}) {\sl independent of the field}. 
A straightforward calculation shows that the perpendicular field
reduces the {\sl renormalized gap} 
by the factor (\ref{new-suppression-factor}) obtained for the plasmon velocity:
\be \label{gap-K-susy}
\tilde \Delta(\phi) = 
{\tilde{\Delta}(0) \over \tilde{g}_0(\phi)} \ ,
\quad 
\tilde \Delta(0) \simeq K^{1/2}(0) D^{1/5} \Delta^{4/5}(0) \ .
\ee
Here $\tilde \Delta(0)$ is the (renormalized) semimetallic gap 
in the absence of the field and $D\simeq \hbar v /R$ is the one-dimensional
bandwidth.

In Eq.~(\ref{gap-K-susy}) 
the value of $K$ is assumed to be taken at $ql_{\rm ch}\sim 1$, where 
$l_{\rm ch}\gg R$ is the size of the charged soliton in the bosonized description.
\cite{Levitov'01}
The universal power law 4/5 in the gap renormalization
(\ref{gap-K-susy}) is valid in the limit $K\gg 1$.
Equations~(\ref{t-v}) and (\ref{gap-K-susy}) show that the characteristic
supersymmetry features of the velocity and minigap suppression 
persist in the presence of Coulomb interactions.
However, the effect of the perpendicular magnetic field is strongly reduced by  
the electron interactions due to the density-of-states increase (\ref{eq:nu}).

%%%%%%%%%%%%%%%%%%%%%%%%%%%%%%%%%%%%%%%%%%%%%%%%%%%%%%%%%%%%%%%%
%%%%%%%%%%%%%%%%%%%%%%%%%%%%%%%%%%%%%%%%%%%%%%%%%%%%%%%%%%%%%%%%
\section{Conclusions}
\label{sec:conclusions}

To conclude, we have shown that the interesting properties of the nanotube electron
spectrum in a perpendicular magnetic field found in Ref.~\cite{Ajiki93JPSJ}
can be understood as a consequence of supersymmetry 
of the low-energy NT Hamiltonian.
%of the Dirac equation on a cylinder. 
We have demonstrated that 
supersymmetry ensures stability of the zero-energy state in metallic NT's
and yields a corresponding enhancement of the density of states.
In semiconducting NT's, supersymmetry leads to an energy gap suppression
that can be observed in transport or tunneling measurements.
We also considered the effects due to the trigonal warping interaction
arising from higher-order gradient expansion terms that violate
supersymmetry and lead to field-sensitive minigaps in the metallic NT spectrum.
Finally, we have found that supersymmetry persists in the presence 
of electron interactions, but the reduction of both 
the renormalized plasmon velocity and the excitation gap 
is weakened due to effectively increased interaction 
strength.

\section*{ACKNOWLEDGMENTS}
We are grateful to Leonid Levitov for inspiring this work and for
numerous suggestions on the manuscript. D.N. acknowledges
discussions with A.~Shytov, D.~Ivanov, M.~Fogler, A.~Vishwanath, and
I.~Gruzberg. H.W.L. was supported by the Nano Research and
Development Program, by the Ministry of Science and Technology in
Korea and Swiss Science Agency in Switzerland through the
Swiss-Korean Outstanding Research Efforts Award program,  and by
the electron Spin Science Center funded by the Korea Science and
Engineering Foundation. The work at MIT was supported by the MRSEC
Program of the National Science Foundation under Grant No. DMR
98-08941.

%%%%%%%%%%%%%%%%%%%%%%%%%%%%%%%%%%%%%%%%%%%%%%%%%%%%%%%%%%%%%%%%

%\end{multicols}

\begin{thebibliography}{99}
\bibitem{k-shift} The intrinsic curvature shifts the Fermi momentum
  slightly and a minigap appears there rather than at the $K$ point.
  Although the quasiperiodic boundary condition does not take into
  account the Fermi momentum shift, it still produces the gap
  and energy dispersion near the gap
  correctly.
\end{thebibliography}
\end{document}